\documentclass[aps,twocolumn,showpacs]{revtex4}
\usepackage{graphicx}
\usepackage{amsmath}
\usepackage{amssymb}

\begin{document}

\title{Towards a Bose-Glass of ultracold atoms in a disordered crystal of light}

\author{L. Fallani, J. E. Lye, V. Guarrera, C. Fort, and M. Inguscio}

\affiliation{LENS European Laboratory for Nonlinear Spectroscopy
and Dipartimento di Fisica, Universit\`a di Firenze\\
via Nello Carrara 1, I-50019 Sesto Fiorentino (FI), Italy}

\begin{abstract}
Starting from one-dimensional Mott Insulators, we use a bichromatic
optical lattice to add controlled disorder to an ideal optical
crystal where bosonic atoms are pinned by repulsive interactions.
Increasing disorder, we observe a broadening of the Mott Insulator
resonances and the transition to an insulating state with a flat
density of excitations, suggesting the formation of a Bose-Glass.
\end{abstract}

\pacs{05.30.Jp, 03.75.Lm, 03.75.Kk, 32.80.Pj, 75.10.Nr, 73.43.Nq}

\maketitle

Disorder plays a crucial role in statistical and condensed matter
physics and it contributes in a substantial way to the mechanism of
transport and conduction. As originally predicted by Anderson
\cite{anderson58}, disorder can lead to localization of a wave
scattered by random impurities. Also interactions are well known to
induce localization effects, as happens in the Mott Insulator (MI)
phase \cite{fisher89,greiner02}, in which a bosonic lattice system
at zero temperature, instead of condensing in a superfluid (SF)
state, forms an insulating ``solid" with integer filling of the
lattice sites. Much theoretical effort has been devoted to
investigate the combined role of disorder and interactions in the
SF-insulator transition observed in many condensed-matter systems
\cite{condmat}. If the disorder in the external potential is large
enough, these systems are expected to enter an insulating state, the
so-called \emph{Bose-Glass} (BG), as predicted in the seminal paper
\cite{fisher89}, characterized by a gapless excitation spectrum and
a finite compressibility
\cite{scalettar91,krauth91,roth03,damski03}.

In this work we add controlled disorder to a collection of 1D
ultracold $^{87}$Rb gases in the MI phase by using a
non-commensurate periodic potential superimposed on the main
lattice, that introduces a randomization of the energy landscape on
the same length scale as the lattice spacing. As a result, the
characteristic resonances in the MI excitation spectrum are lost and
the system rearranges to form a state with vanishing long-range
coherence and a broadband excitability. These observations
constitute the first significative evidence in the direction of
demonstrating a BG phase of ultracold atoms.

At zero temperature the many-body quantum state of an interacting
gas of identical bosons in a lattice potential is well described by
the Bose-Hubbard Hamiltonian  $\hat{H} = -J \sum_{\left<j,j'\right>}
\hat{b}_j^\dagger \hat{b}_{j'} + \frac{U}{2} \sum_{j} \hat{n}_j
\left( \hat{n}_j - 1 \right) + \sum_{j} \epsilon_j \hat{n}_j$, where
$\hat{b}_j$ ($\hat{b}_j^\dagger$) is the destruction (creation)
operator of one particle in the $j$-th site,
$\hat{n}_j=\hat{b}_j^\dagger \hat{b}_j$ is the number operator, and
$\left<j,j'\right>$ indicates the sum on nearest neighbors
\cite{fisher89,jaksch98}. The total energy results from the sum of
three terms: $J$ is the \emph{hopping energy}, proportional to the
probability of tunnelling between adjacent sites, $U$ is the
\emph{interaction energy}, arising from on-site interactions
(repulsive for $^{87}$Rb, for which $U>0$), and $\epsilon_j$ is a
site-dependent energy accounting for inhomogeneous external
potentials.

\begin{figure}[b!]
\begin{center}
\includegraphics[width=0.80\columnwidth]{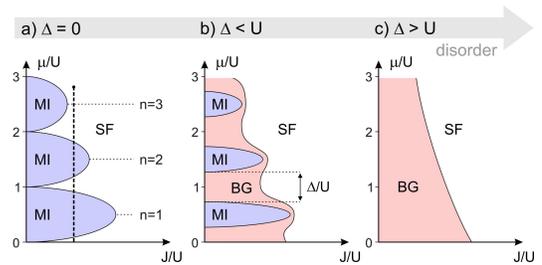}
\end{center}
\caption{Phase diagram for disordered interacting bosons
\cite{fisher89}. Depending on the ratio between tunnelling energy
$J$, interaction energy $U$ and disorder $\Delta$, the system forms
a superfluid (SF), a Mott-Insulator (MI) or a Bose-Glass (BG)
state.} \label{fig:diagramma}
\end{figure}

When $\epsilon_j=0$ the ground state is determined by the
competition between $J$ and $U$. When $U \ll J$ the system is in a
SF state, in which the bosons are delocalized and tunnelling ensures
long-range coherence. Instead, when $U \gg J$, the system is in a
localized MI state, where phase coherence is lost and number Fock
states are created at the lattice sites. The transition from a SF to
a MI for ultracold atoms in an optical lattice has been first
reported for a 3D system in \cite{greiner02} and for an array of 1D
gases in \cite{stoferle04}. The phase diagram of the system depends
on the chemical potential $\mu$ (related to the atomic density) and
shows the existence of MI lobes with integer number of atoms per
site (Fig. \ref{fig:diagramma}a). In the experiments, an additional
harmonic confinement is present, resulting in a smooth variation of
the density across the sample. As a result, in a single experimental
run one typically averages on an extended range of local $\mu$
(vertical dotted line in Fig. \ref{fig:diagramma}a). This
inhomogeneity precludes a sharp phase transition, due to the
coexistence of SF and MI near the critical point, and is responsible
for the formation of domains with different fillings deep in the MI
phase \cite{batrouni02}.

In the presence of a disordered external potential an additional
energy scale $\Delta$ enters the description of the system. We
consider the case of bounded disorder, in which $\epsilon_j \in
[-\Delta/2,\Delta/2]$. In the presence of weak disorder $\Delta<U$
the MI lobes should progressively shrink and a new BG phase should
appear (Fig. \ref{fig:diagramma}b), eventually washing away the MI
region for $\Delta>U$ (Fig. \ref{fig:diagramma}c) \cite{fisher89}.
This BG phase shares some properties with the MI state, namely both
are insulating states, with vanishing long-range coherence and
vanishing superfluid fraction. However, differently from the MI, the
BG presents a gapless excitation spectrum and a finite
compressibility.

\begin{figure}[t!]
\begin{center}
\includegraphics[width=0.70\columnwidth]{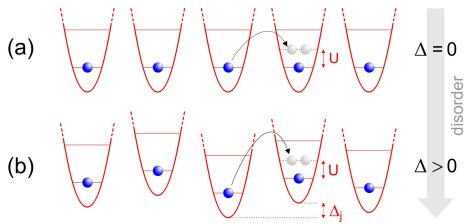}
\end{center}
\caption{In a homogeneous MI the tunnelling of one boson from a site
to a neighboring one has an energy cost $\Delta E=U$. In the
presence of disorder this excitation energy becomes $\Delta E=U \pm
\Delta_j$, where $\Delta_j$ is the local site-to-site energy
difference.} \label{fig:excitation}
\end{figure}

In order to understand the physics happening when approaching the BG
phase, we consider the limit $J \to 0$ and unitary filling of the
lattice sites. In a MI an energy gap exists, since the elementary
excitation - the hopping of a boson from a site to a neighboring one
- has an energy cost $U$. The presence of disorder introduces random
energy differences $\Delta_j \in [-\Delta,\Delta]$ between
neighboring sites (see Fig. \ref{fig:excitation}b) and the energy
cost for such a process becomes $U \pm \Delta_j$, that depends on
the position. In the full BG, when $\Delta \gtrsim U$, an infinite
system can be excited at arbitrarily small energies and the energy
gap shrinks to zero. Despite the zero energy gap, excitations only
occur locally and the BG remains globally insulating.

For a system of ultracold bosons in an optical lattice one can
introduce controlled disorder in different ways. Experiments have
already been realized with laser speckles in combination with atomic
Bose-Einstein condensates (BECs)
\cite{lye05,fort05,clement05,schulte05}. Optical disorder on much
smaller length scales can be obtained by using two-color lattices
\cite{roth03,damski03}, i.e. superimposing on the already existing
lattice a second weaker lattice with non-commensurate spacing.

\begin{figure}[b!]
\begin{center}
\includegraphics[width=\columnwidth]{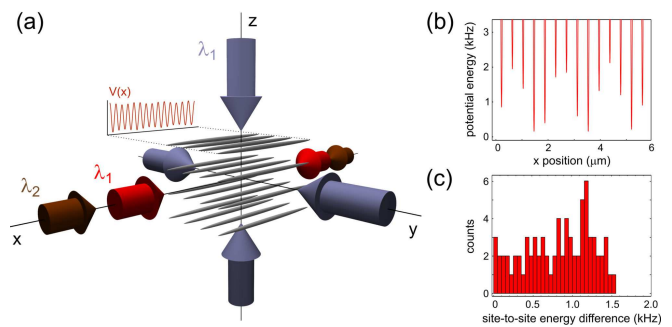}
\end{center}
\caption{a) A collection of 1D Bose gases in a deep 2D lattice
experience the bichromatic potential $V(x)$. b) Potential energy
along $V(x)$ for $s_1=16$ and $s_2=1$. c) Histogram of the energy
differences between neighboring sites for $s_2=1$.}
\label{fig:setup}
\end{figure}

In the experiment, an optical lattice (\emph{main lattice}) is
produced by using a Titanium:Sapphire laser at $\lambda_1=830$ nm.
Disorder is introduced by using an auxiliary lattice
(\emph{disordering lattice}) obtained from a fiber-amplified diode
laser at $\lambda_2=1076$ nm. The resulting potential along the
lattice axis $\hat{x}$ is $V(x)=s_1 E_{R1} \sin^2 (k_1x) + s_2
E_{R2} \sin^2 (k_2 x)$, where $s_1$ and $s_2$ measure the height of
the lattices in units of the recoil energies
$E_{R1}=h^2/(2m\lambda_1^2)\simeq h \times 3.33$ kHz and
$E_{R2}=h^2/(2m\lambda_2^2)\simeq h \times 1.98$ kHz, $h$ is the
Planck constant and $m$ the mass of a $^{87}$Rb atom. When $s_2 \ll
s_1$ the disordering lattice has the only effect to scramble the
energies $\epsilon_j$, which are non-periodically modulated at the
length scale of the beating between the two lattices $(2/\lambda_1
-2/\lambda_2 )^{-1} = 1.8$ $\mu$m, corresponding to 4.3 lattice
sites (see Fig. \ref{fig:setup}b). In Fig. \ref{fig:setup}c we plot
a histogram of the site-to-site energy differences
$|\Delta_j|=|\epsilon_j-\epsilon_{j-1}|$ occurring over the size of
our samples (32 $\mu$m) for $s_2=1$. Recent theoretical works
\cite{roth03,damski03,hild06} have demonstrated that in finite-sized
systems this quasiperiodic potential can produce the same effects
induced by a truly random potential and allow the observation of a
BG of ultracold atoms.

We perform the experiment with a collection of $\approx 10^3$
independent 1D bosonic systems tightly confined in the sites of a 2D
optical lattice (\emph{trapping lattice}) produced with laser light
at $\lambda_1=830$ nm and aligned along $\hat{y}$ and $\hat{z}$ axes
(see Fig. \ref{fig:setup}a). The source of ultracold atoms is
provided by a BEC of $1.5 \times 10^5$ $^{87}$Rb atoms. In order to
create the atom tubes we adiabatically increase the intensity of the
2D trapping lattice by using a 100 ms long exponential ramp with
time constant 30 ms. After the ramp the lattice height, measured in
units of $E_{R1}$, is $s_\perp=40$ and the atoms are confined in the
lattice sites with trapping frequency $\nu_\perp=42$ kHz along
$\hat{y}$ and $\hat{z}$. The confinement along the tubes is much
more loose, being caused by the gaussian shape of the laser beams
and the harmonic magnetic trap, resulting in a trapping frequency
$\nu_x=75$ Hz. The single-particle tunnelling rate between adjacent
tubes is $J_y=J_z=0.4$ Hz, that can be completely neglected on the
timescale of the experiments.

Together with the 2D lattice we switch on the bichromatic lattice
at wavelengths $\lambda_1$ and $\lambda_2$ along the direction of
the tubes (see Fig. \ref{fig:setup}a) by using the same
exponential ramp. Then we characterize the many-body state of the
1D systems by measuring the excitation spectrum and observing the
interference pattern after time-of-flight (TOF), i.e. switching
off the confining potentials and imaging the density distribution
after expansion.

The excitation spectrum of the 1D gases in the bichromatic lattice
is measured with the Bragg spectroscopy technique introduced in
\cite{stoferle04}. A sinusoidal modulation of the main lattice
height $s_1$ with frequency $\nu$ and amplitude $30\%$ stimulates
the resonant production of excitations with energy $h\nu$. We detect
the excitations produced after 30 ms of modulation by first
decreasing in 15 ms the intensity of the lattices back to
$s_1=s_\perp=5$, $s_2=0$ in the 3D SF phase, waiting 5 ms and then
switching off the potentials and imaging the atoms after TOF. The
width of the central peak in the images is related to the energy
transferred to the atomic system, giving information on the
excitability of the sample at that frequency \cite{stoferle04}.

\begin{figure}[t!]
\begin{center}
\includegraphics[width=0.85\columnwidth]{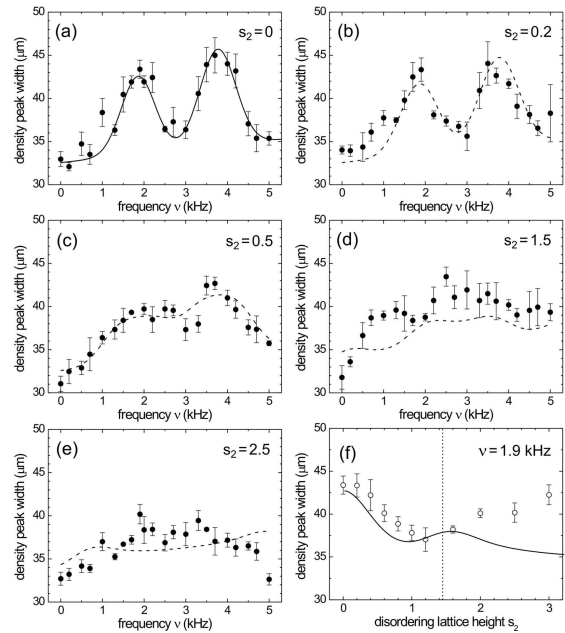}
\end{center}
\caption{a-e) Excitation spectra for $s_1=16$ and different values
of $s_2$. f) Effect of the modulation at frequency $\nu=1.9$ kHz as
a function of $s_2$. The line in a) is a fit to the experimental
points, while the curves in b-f) are calculated from a model of
inhomogeneous broadening of the MI resonances.} \label{fig:spettri}
\end{figure}

A typical spectrum for the MI state with $s_1=16$ and $s_2=0$ is
shown in Fig. \ref{fig:spettri}a. Here we detect the presence of an
excitation peak at $\nu=1.9(1)$ kHz, corresponding to the
interaction energy $U$. At twice the frequency 3.8(1) kHz we observe
a second peak \cite{greiner02}, that can be attributed to
higher-order processes and to excitations taking place at the
boundary between different MI domains. For these parameters the peak
chemical potential is $\mu \simeq 2.8U$ (see vertical line in Fig.
\ref{fig:diagramma}), giving rise to domains with 1, 2 and 3 atoms
per site. The same measurements are then repeated in the presence of
the disordering lattice. Fig. \ref{fig:spettri}b-e show the measured
spectra for increasing disorder from $s_2=0.2$ to $s_2=2.5$. At $s_2
= 0.5$ (Fig. \ref{fig:spettri}c) one already detects the
disappearance of the characteristic peak structure of the MI and the
appearance of a broader spectrum. At the largest disorder height
$s_2=2.5$, when the maximum energy difference $\Delta$ between
neighboring sites is $3.2$ kHz $\simeq 1.7U$ and one expects to have
entered the full BG phase, the resonances are completely lost.

\begin{figure}[t!]
\begin{center}
\includegraphics[width=0.95\columnwidth]{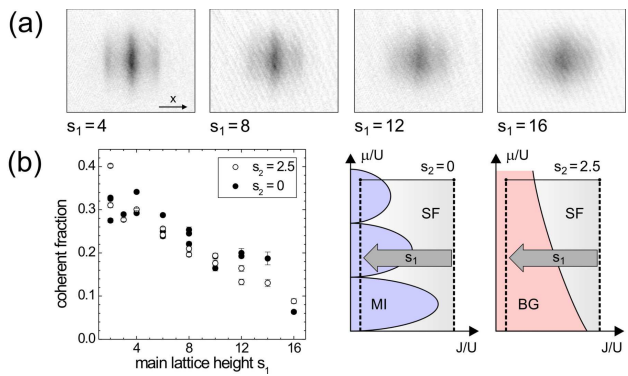}
\end{center}
\caption{a) Density distribution after time-of-flight for different
values of $s_1$ and $s_2=2.5$. b) Coherent fraction as a function of
$s_1$ for $s_2=0$ and $s_2=2.5$ and sketch of the corresponding
scans in the phase diagrams. Each point corresponds to a different
set of images ($\approx 10$ images per set).} \label{fig:trans}
\end{figure}

Additional information can be obtained by analyzing the TOF images,
that provide a measurement of phase coherence. We first prepare the
system in a state with arbitrary $s_1$ and $s_2$, then we suddenly
ramp in $\approx 40$ $\mu$s the lattice heights to $s_1=25$ and
$s_2=0$ (while keeping $s_\perp=40$) and finally switch off the
confining potentials. This time is short enough not to change the
coherence properties of the system, but allows us to project the
state under investigation onto a same reference state
\cite{stoferle04}. In Fig. \ref{fig:trans}a we report images of the
density distribution after a TOF of 20 ms for $s_2=2.5$ and
different values of $s_1$. The presence of vertical interference
fringes is an indicator of long-range coherence along the tubes.
When increasing the height of the main lattice, we observe a
progressive loss of coherence indicating the transition from a SF to
an insulating state \cite{notacoerenza}. In Fig. \ref{fig:trans}b we
report the coherent fraction, measured as the number of atoms in the
interference peaks divided by the total number of atoms
\cite{stoferle04} both for $s_2=0$ and for $s_2=2.5$.

The combination of the excitation spectra measurements and the TOF
images indicates that, with increasing disorder, the system goes
from a MI to a state with vanishing long range coherence and a flat
density of excitations. The concurrence of these two properties
cannot be found in either a SF or an ordered MI, and is consistent
with the formation of a BG. Rigorously speaking, the BG phase should
be characterized by a gapless spectrum. Detecting the absence of a
gap is technically challenging, since it would require a measurement
of excitability at arbitrarily small energies. A direct measurement
of a small energy gap cannot be accomplished with the modulation
technique we have used, that works well only for frequencies $\nu$
much larger than the reciprocal of the modulation time $\tau=30$ ms,
i.e. starting from a few hundred Hz. However, the excitation
spectrum is expected to be gapless only for an infinite system,
while finite-sized systems always have discrete energy spectra.
Nevertheless, we expect the density of excitations in a finite-sized
BG to lose the characteristic resonances of the MI and to become
flat \cite{fisher89}. In the following we will show that for weak
disorder the spectra in Fig. \ref{fig:spettri} can be explained with
the inhomogeneous broadening of the MI peaks, accompanied by the
consequent reduction of the gap, which is the first prerequisite for
the formation of a BG.

We have developed a model in which we calculate the inhomogeneous
broadening of the MI resonances at $U$ and $2U$ caused by the
disordered distribution of site-to-site energy differences
$\Delta_j$ (see Fig. \ref{fig:excitation}b). The broadened spectra
are obtained by convolving the MI spectrum (fitted in Fig.
\ref{fig:spettri}a with a double gaussian + a linear pedestal, solid
line) with the distribution of energy shifts across the lattice. The
result of this convolution is reported in the dashed curves of Fig.
\ref{fig:spettri}b-e, showing a fine agreement with the experimental
data for $s_2 \leqslant 1$ \cite{workinprogress}.

In order to more quantitatively analyze the disappearance of the MI
resonances, we report in Fig. \ref{fig:spettri}f the width of the
density distribution after an excitation at $\nu=1.9$ kHz,
corresponding to the first resonance in the MI spectrum, as a
function of $s_2$ (empty circles). The solid line is the theoretical
prediction based on the model of inhomogeneous broadening, that
nicely match the experimental findings for $s_2<1.5$. One expects
this broadening to happen when approaching the transition from a MI
to a BG, with the energy gap progressively closing with increasing
disorder \cite{krauth91}. Eventually, when the broadened resonances
reach zero-energy and the gap completely disappears, the transition
to a BG is expected to occur. We find that the agreement with the
model breaks down for $s_2>1.45$ (vertical dotted line), when indeed
the maximum site-to-site energy difference is larger than $U$. At
this point the atoms rearrange in a new insulating state with
different sites filling, and the simple model of MI inhomogeneous
broadening breaks down.

Similar results for the excitation spectrum have been recently
predicted in \cite{hild06}, where the authors study the response of
a 1D Bose gas to a periodic amplitude modulation of a superlattice.
Our experimental findings are in good agreement with the numerical
results of \cite{hild06}, showing a broadening of the MI resonances
for weak disorder and their complete disappearance when entering the
BG.

Despite this remarkable change in the excitation spectrum, an
undeniable hallmark of the formation of a BG is still lacking. In
particular, if a precise detection of the vanishing gap could be
performed, it would discriminate between a genuine BG and a strongly
inhomogeneous MI. To this aim a measurement of compressibility would
provide a decisive proof for the creation of such a state. However,
in actual MI experiments the existence of multiple domains leads
itself to already globally compressible systems \cite{batrouni02}.
Regarding the coherence properties, few theoretical works predict
that for restricted ranges of parameters one should observe an
enhancement of coherence due to the presence of disorder in the
route to the BG \cite{sengupta06}. As shown in Fig.
\ref{fig:trans}b, we do not observe any appreciable difference
varying $s_2$ within the reproducibility limits of our experiment
\cite{nota1}. However, a quantitative prediction of these effects
for real experimental systems is absent. Furthermore, in the
progression from SF to MI/BG, one simultaneously probes extended
regions of the phase diagram (as sketched in Fig. \ref{fig:trans}b),
that smooth the transition and could smear out these effects.

In conclusion, we have added controlled disorder onto a collection
of 1D bosonic gases in an optical lattice by means of an additional
lattice with different spacing. We have reported on the first
observation of the transition from a Mott Insulator to a state with
vanishing coherence and a flat density of excitations. These
coexisting properties suggest the formation of a Bose-Glass, which
is expected to appear for our parameters \cite{roth03, damski03,
hild06}. Future work will be done in the direction of implementing
new techniques for a more exhaustive characterization of this novel
state, possibly including a measurement of compressibility and new
detection schemes \cite{folling05}.

This work has been funded by the EU Contract No. HPRN-CT-2000-00125,
MIUR FIRB 2001, MIUR PRIN 2005 and Ente Cassa di Risparmio di
Firenze. We acknowledge E.~A.~Cornell, P.~Zoller, R.~Fazio,
C.~Tozzo, F.~S.~Cataliotti, M.~Modugno, D.~S.~Wiersma and all the
Cold Quantum Gases Group at LENS for stimulating comments. We thank
P. Cancio Pastor and P. De Natale from INOA for kind concession of
the infrared laser.


\begin{thebibliography}{99}
\bibitem{notacoerenza}
The technique used for the excitation spectrum measurements ensures
itself that this loss of coherence is reversible. Indeed, when
coming back to $s_1=s_\perp=5$, $s_2=0$, the peaks in the TOF images
indicate that coherence can be reestablished, even for $s_2 \neq 0$.
\bibitem{workinprogress}
More details will be presented in a future publication.
\bibitem{nota1}
We have also considered different values of $s_2$ ranging from 0.8
to 4 obtaining similar results. We have also measured the coherence
length (by taking the width of the peaks), still finding no
appreciable differences varying $s_2$.
\end{thebibliography}
\end{document}